\documentclass[%
 reprint,
 amsmath,amssymb,
 aps,
]{revtex4-1}
\usepackage{color}
\usepackage{graphicx}% Include figure files
\usepackage{dcolumn}% Align table columns on decimal point
\usepackage{float}
\usepackage{bm}% bold math
\usepackage{inputenc}

\begin{document}

\preprint{APS/123-QED}

\title{Electric Field Driven Memristive Behavior at the Schottky Interface of Nb- doped SrTiO$_3$}% Force line breaks with \\
%\thanks{A footnote to the article title}%

\author{A. S. Goossens}
\thanks{corresponding author}
\email{a.s.goossens@student.rug.nl}
\author{A. Das}
\author{T. Banerjee}
\thanks{corresponding author}
\email{t.banerjee@rug.nl}
\affiliation{University of Groningen, Zernike Institute for Advanced Materials, 9747 AG Groningen, The Netherlands}

\date{\today}% It is always \today, today,
             %  but any date may be explicitly specified

\begin{abstract}
Computing inspired by the human brain requires a massive parallel architecture of low-power consuming elements of which the internal state can be changed. SrTiO$_3$ is a complex oxide that offers rich electronic properties; here Schottky contacts on Nb-doped SrTiO$_3$ are demonstrated as memristive elements for neuromorphic computing. The electric field at the Schottky interface alters the conductivity of these devices in an analog fashion, which is important for mimicking synaptic plasticity. Promising power consumption and endurance characteristics are observed. The resistance states are shown to emulate the forgetting process of the brain. A charge trapping model is proposed to explain the switching behavior.
\end{abstract}

\pacs{Valid PACS appear here}% PACS, the Physics and Astronomy
                             % Classification Scheme.
%\keywords{Suggested keywords}%Use showkeys class option if keyword
                              %display desired
\maketitle

%\tableofcontents

\section{\label{sec:level1}Introduction}
Modern day computers rely heavily on the \textit{von Neumann} \cite{VonNeumann1945} architecture and, while for many tasks this computing approach works well, it cannot compete with the speed, low power consumption and adaptability of the human brain for certain functions, such as performing cognitive tasks. This has led to a need for alternative architectures inspired by the performance of the brain \cite{Yang2013,Schuman2017,Covi2016,Grollier,Indiveri2015}. In such neuromorphic architectures, memory and processing operations are ideally co-located in devices that are highly connected, capable of computing in parallel yielding an analog output while consuming low power \cite{Yang2013,Schuman2017,Grollier,Indiveri2015,Kuzum2012}. In order to mimic such performance of the brain, multi-level storage devices and a high storage density are needed \cite{Covi2016,Boyn2017,Indiveri2015,Kuzum2012,Yang2012}. The exact required device performance specifications, including switching speed and retention, strongly depend on the biological functionality that is required from the neural network \cite{Schuman2017,Yang2013,Grollier,Indiveri2015}. For example, the processes of learning and forgetting in the human brain are governed by the modulation of the strengths of paths connecting neurons, known as synaptic weights, as a result of action potentials \cite{Abbas2018,Covi2016,Yin2017,Kuzum2012}. This \textit{synaptic plasticity} behavior can be emulated by memristors \cite{Chua1971}: two-terminal devices of which the microscopic internal state can be modified and measured as a change in the conductance. Consequently, these devices can be used to emulate synapses for neuromorphic computing \cite{Abbas2018,Boyn2017,Yin2017,Covi2016}. These memristors should have sizes on the order of nanometers and should offer low power consuming resistive switching in order to be able to build a highly connected parallel architecture \cite{Grollier}. These features make these devices very computationally efficient, as well as fault tolerant; connections that do not work can be bypassed through alternative paths.
\\
\indent
In many material systems where memristive behavior has been demonstrated, resistive switching is absent in the as-fabricated device and an electroforming process, during which the device is subjected to a high bias, is needed to induce the switching \cite{Abbas2018,JoshuaYang2009,Yang2013,Sawa2008}. The details of the forming process are well discussed in literature, but pose limitation on the operation of memristors at low powers due to the requirement of the high forming voltage. Additionally, a lack of control over the forming process can increase device-to-device variations and reduce device performance \cite{Kim2016,Yang2013,Yang2012,Abbas2018}. In this context, interface driven memristive effects such as that found in Schottky interfaces between Nb-doped SrTiO$_3$ (Nb:STO) and metals, without the need of a forming process, are an important consideration \cite{Mikheev2014,Muenstermann,Yin2015,Yin2017,Dong-Wook2010,Chen2011,Park2008,Seong2007,Ni2007,Sawa2008}. Recently, there has been a growing interest in understanding the rich and intricate properties of complex oxide interfaces and their application in superconductivity, magnetism and multiferroics \cite{Hwang2012,BenShalom2009,Caviglia2010,Sulpizio2014}. The Schottky interface of Nb:STO offers rich electronic properties that have been shown to be useful both for memristor and spintronics applications \cite{Kamerbeek2015,Han2013,Kamerbeek2018,Das2018}. An important parameter is the non-linear variation of the large dielectric permittivity ($\epsilon_r$) with temperature and electric field \cite{Neville1972}. The tuning of the potential profile of the conduction band of Nb:STO with electric field and temperature gives rise to charge transport characteristics that are unique and are not observed in conventional semiconductors like Si, GaAs and Ge.
\begin{figure*}[]
\centering
\includegraphics[scale=0.6]{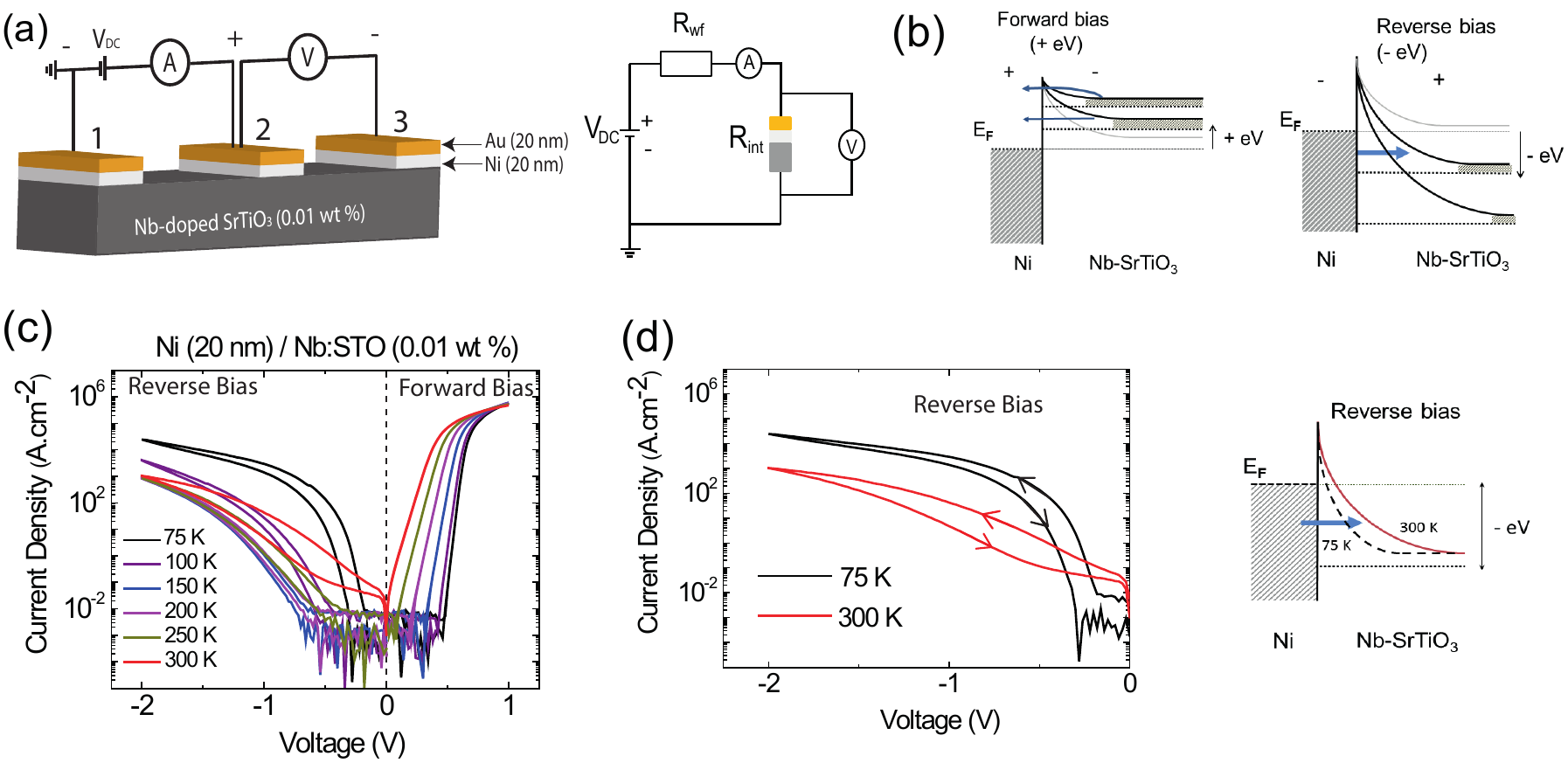}
\caption{(a) Three terminal (3T) device geometry where a voltage (V$_{DC}$) is sourced  between central contact 2 and a grounded reference contact 1, giving rise to a potential drop between contact 2  and reference contact 3. The measurement scheme used is 4-probe and the junction voltage (resistance R$_{int}$) is measured, decoupled from the resistances due to the leads and the semiconductor channel ($R_{wf}$). (b) Potential profiles for different polarities of V$_{DC}$: positive (forward bias) and negative (reverse bias) across the Schottky interface of Nb:STO with Ni at room temperature. (c) Temperature dependent I-V characteristics for a contact (100 $\times$ 200 $\mu$m$^2$) where the current density is plotted with junction voltage. (d) Increase in the reverse bias current on decreasing temperature, as shown by the crossover of the reverse current between 75 K and 300 K. The potential profile shows narrowing of the conduction band in STO at 75 K resulting in increasing current due to enhanced tunneling. 
}
\label{fig:fig1}
\end{figure*}
\\ \indent
In this manuscript, the use of Ni/Nb:SrTiO$_3$ Schottky junctions for potential applications in neuromorphic memristors is evaluated. Here, relatively large structures for Ni contact pillars (100 $\times$ 200 $\mu$m$^2$) on Nb:STO are used, the simplistic structure of these pillars, however, lends itself well to the required down-scaling. In the first part of this work, the possibility of multi-level switching is demonstrated by gradually sweeping the applied voltage. By changing the magnitude of the reverse bias \textit{reset} voltage, the devices can be switched to higher resistance states so that multi-level storage can be realized. From this, important information about the resistance switching mechanisms is also inferred. The second part of this work focused on how the resistance state can be changed by applying voltage pulses. This is of importance for emulating synaptic behavior, where resistance changes are brought about by the spiking behavior of neurons \cite{Yin2017}. The reliability of the devices, with respect to retention and endurance, was investigated by performing repeat cycles of resistance switching.
\section{\label{sec:experiments}Experimental Details and Results}
Devices were fabricated on n-type doped (001) SrTiO$_3$ single crystalline substrates with 0.01 wt$\%$ Nb doping, obtained from Crystec GmbH. To remove the SrO sublattice from the top surface and obtain TiO$_2$ termination the substrates were treated with buffered hydrofluoric acid and deionized water. This was followed by the immediate deposition of 20 nm of Ni, capped with 20 nm of Au by electron beam evaporation at a base pressure of 10 $^{-6}$ Torr. Three terminal (3T) geometry meso-structures were fabricated by UV lithography and Ar-ion etching. A typical device geometry is shown in Fig. \ref{fig:fig1}(a). \\
\begin{figure*}[t]
    \centering
    \includegraphics[width=\textwidth]{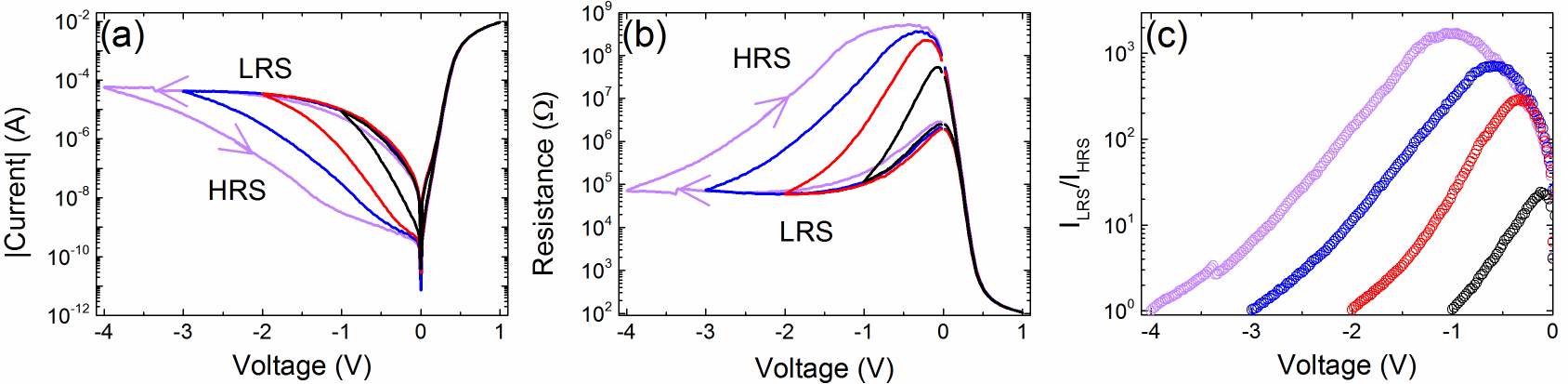}
    \caption{(a) Room temperature I-V characteristics of the device obtained by performing repeat cycles in which the voltage was swept between +1 V and a varying (-1 V (black), -2 V (red), -3 V (blue) and -4 V (purple)) reverse bias voltage. Each subsequent cycle was done after a delay time of 300 s at +1 V. The upper branch of each cycle is the low resistance state (LRS) and the lower branches are the high resistance states (HRSs). (b) Resistances calculated from the cycles performed in (a); the lower branches are the trace directions and the upper branches are the retrace directions. (c) Ratios of the HRS (lower branch) and LRS (upper branch) resistance values calculated from (a).}
    \label{fig:fig2}
\end{figure*}
%\indent
In order to investigate the Schottky contact of Ni on Nb:STO, current-voltage (I-V) measurements were performed in a 4-probe, three terminal (3T) geometry using a Keithley 2410. In this geometry, a dc-voltage ($V_{DC}$) is sourced between a central contact 2 and a grounded reference contact 1 (shown in Fig. \ref{fig:fig1}(a)), giving rise to a potential drop between contact 2 and reference contact 3. The dc-current is measured between contact 2 and source contact 1. This geometry ensures that the potential drop ($V$) is caused solely by the interface resistance ($R_{int}$) of contact 2 and eliminates series resistances from the leads, semiconducting channel ($R_{wf}$) and the two reference contacts. Isolating the effect of a single Schottky interface gives a better understanding of the transport mechanisms that are controlled by the electric field across the interface with Nb:STO. The temperature-dependent I-V characteristics of a junction of 100 $\times$ 200 $\mu$m$^2$ are shown in Fig. \ref{fig:fig1}(c). At each temperature, the current density is plotted with respect to V, following a sweep from 0 V$\rightarrow$ +1 V$\rightarrow$ 0 V $\rightarrow$ -2 V$\rightarrow$ 0 V. The I-V characteristics clearly show the existence of a Schottky barrier at the Ni/Nb:STO interface. Transport in forward bias is governed by thermionic emission. Upon increasing the voltage, the conduction band of Nb:STO is raised with respect to the Fermi level ($E_F$) of Ni, as shown in Fig.\ref{fig:fig1}(b), resulting in a decrease in the width of the depletion region and the built-in electric field. At low temperatures, there is less thermal energy available for electrons to overcome the barrier and consequently larger voltages are required for transport compared to higher temperatures. Current density, $J$, can be described by:
\begin{equation}
    J(V)=A^*T^2e^{-\frac{q\phi_B}{k_BT}}\Big(e^{\frac{qV}{nk_BT}}-1\Big)
    \label{Eq:thermionic}
\end{equation}
where $A^*$ is the Richardson constant, $q$ the electron charge, $\phi_B$ the Schottky barrier height, $k_B$ the Boltzmann constant, $n$ the ideality factor, $V$ the applied voltage and $T$ the temperature.
\begin{figure}[]
    \centering
    \includegraphics[scale=0.35]{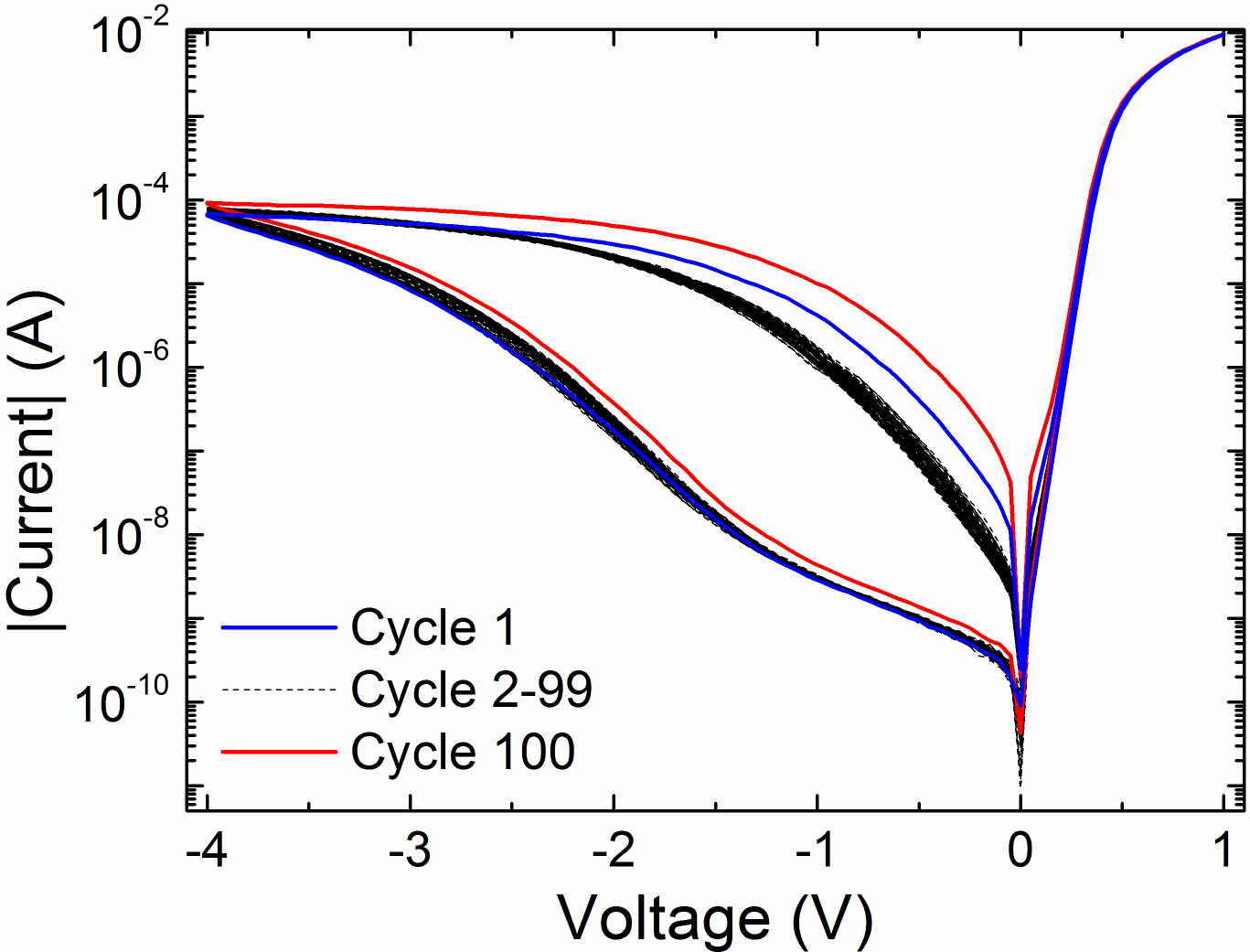}
    \caption{Endurance characteristics of the device over 100 cycles subsequent switching cycles. The First (blue) to 99$^{th}$ cycle were performed following a sequence where, after a 120 s wait time at +1 V, the voltage was swept from +1 V to -4 V and back to +1 V. The black lines show cycle 2-99. Cycle 100 (red) was obtained after a 1 hour \textit{set} time at +1 V.}
    \label{fig:fig3}
\end{figure}
\\ \indent
In reverse bias (negative V$_{DC}$), a strong response of the dielectric permittivity of STO ($\epsilon_r$) to the increasing built-in electric field results in a steeper bending of the conduction band of STO as shown in Fig. \ref{fig:fig1}(b). This decrease in the effective barrier width at $E_F$ promotes tunneling, allowing current to flow in the reverse bias regime. The reverse current increases upon lowering the temperature as illustrated in Fig. \ref{fig:fig1}(d) where 75 K and 300 K are compared. Upon reducing temperature, $\epsilon_r$ increases and becomes more sensitive to electric fields, resulting in a significant narrowing of the conduction band potential and allowing for enhanced tunneling \cite{Susaki2007,Kamerbeek2014}. In summary, the charge transport is governed by thermionic emission of charge carriers in the forward bias regime and by temperature independent field emission in reverse bias. \\
\indent
%\FloatBarrier
The reverse bias hysteretic response of the resistance embodies the colossal electroresistance (CER) effect. The Schottky interface of Nb:STO is replete with trapped charge states in the form of oxygen vacancies that occupy energy levels about 0.5 to 1 eV below the conduction band \cite{Andra,Mitra2012,Saraf2016}. Their accessibility to electrons flowing into the conduction band in Nb:STO is enhanced by the increasing electric field, giving rise to the CER effect. This switching is bipolar: the resistance will increase in reverse bias and decrease in forward bias. 
\\
\indent
To demonstrate multi-level resistance switching, the bias voltage was swept from +1 V $\rightarrow$ -V$_{reset}$ $\rightarrow$ +1 V. The \textit{set} voltage (+1 V) and time (300 s), to switch to a low resistance state (LRS), was kept constant while the \textit{reset} voltage (V$_{reset}$) was varied (-1 V, -2 V, -3 V and -4 V) to \textit{reset} to different high resistance states (HRSs). The results of this are shown in Fig. \ref{fig:fig2}(a) and illustrate that by changing the \textit{reset} voltage, states of increasingly higher resistance can be attained. The resistance values corresponding to every part of the loops are explicitly plotted in Fig. \ref{fig:fig2}(b). The resistance of the LRS does not change significantly by changing the bias, while the resistance of the HRS peaks at a low bias value and thereafter decreases until reaching the resistance value of the LRS at the maximum \textit{reset} voltage. This is mimicked in Fig. \ref{fig:fig2}(c), where the ratio of the resistance of the HRS over the LRS is extracted at every reverse bias value.
\begin{figure*}[]
    \centering
    \includegraphics[width=\textwidth]{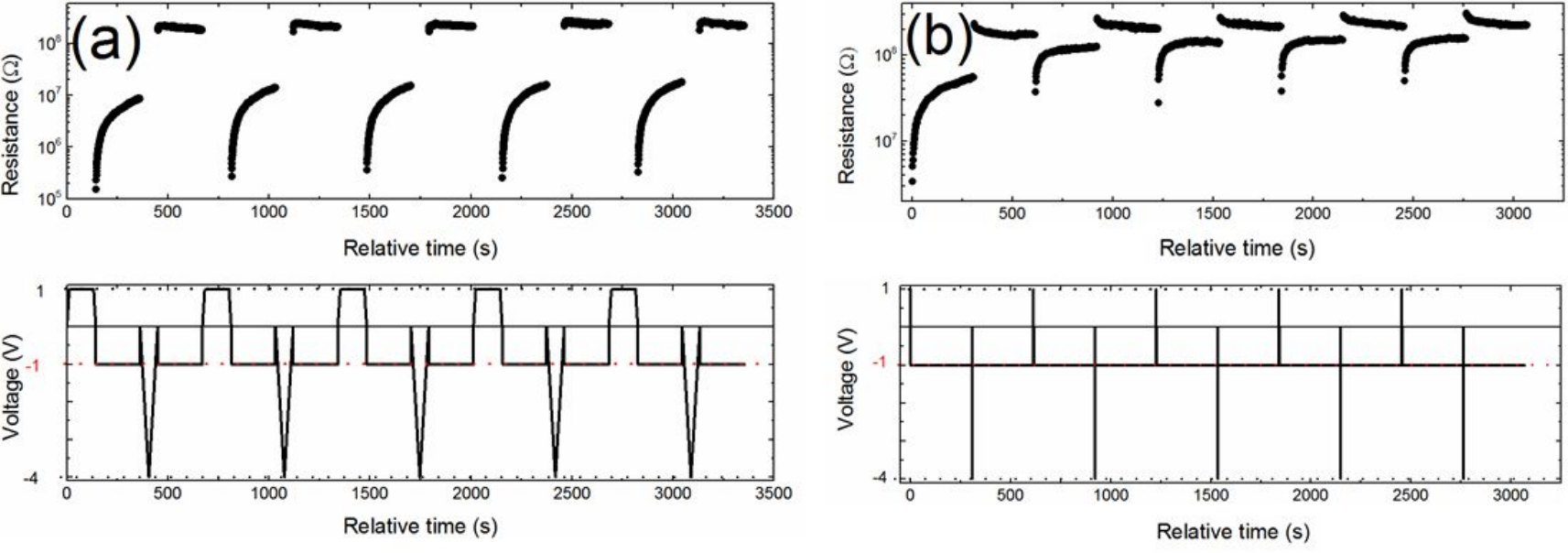}
    \caption{Retention characteristics of the LRS and HRS upon repeatedly switching between the two states by (a) voltage sweeps and (b) voltage pulses.}
    \label{fig:fig4}
\end{figure*}
\\ \indent
The room temperature endurance and cycle-to-cycle resistance variations were understood by performing 100 subsequent cycles between +1 V and -4 V and back, shown in Fig. \ref{fig:fig3}. The first 99 cycles were done with a \textit{set} time of 120 s at +1 V between cycles. The last cycle was performed after a 1 hour \textit{set} time at +1 V. There is minimal variation in the HRS between cycles, while the resistance of the LRS increases upon cycling. This indicates that the 120 s used to \textit{set} the device is insufficiently long to completely restore the LRS. These data show that the device is able to repeatedly switch its resistive state over at least 100 consecutive cycles without showing any degradation in device performance.
\\
\indent
The retention of the LRS and HRS was investigated by setting the respective states in two ways. Firstly, by the voltage sequence shown in the bottom panel of Fig. \ref{fig:fig4}(a): here the LRS was \textit{set} by sweeping 0 V $\rightarrow$ +1 V$\rightarrow$ 0 V with a \textit{set} time of 120 s at +1 V. This was followed by a negative bias read pulse of -1 V. The device was subsequently \textit{set} to a HRS by sweeping 0 V $\rightarrow$ -4 V $\rightarrow$ 0 V and another read pulse was applied. Secondly, the LRS and HRS were \textit{set} by applying short pulses of 0.9 s of +1 V and -4 V respectively, depicted in Fig. \ref{fig:fig4}(b). In both cases, the sequences were repeated five times. 
\\ \indent The two sets of results demonstrate that the resistance of the HRS does not significantly change either over time or between cycles. The differences in the LRSs resulting from the two setting methods on the other hand, clearly show deviations. When a short pulse is used to \textit{set} the LRS, the resulting resistance difference between the LRS and HRS is smaller than when a longer pulse is used to \textit{set} this state. Over time, the resistance of the LRS becomes higher and for the short pulse setting method the high and low resistance states reach similar resistance values more rapidly.
\section{Discussion}
\begin{figure}[H]
    \centering
    \includegraphics[scale=0.3]{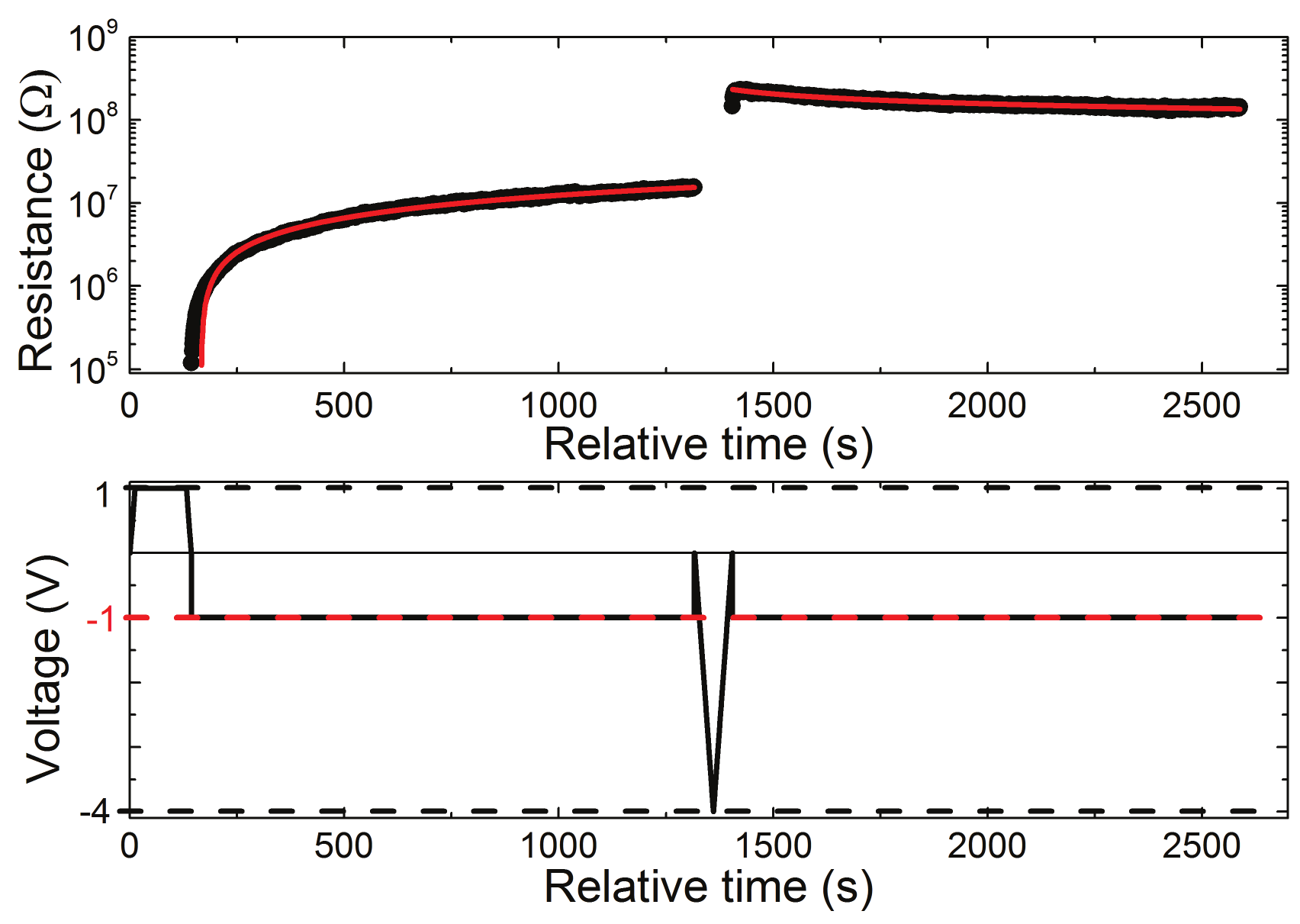}
    \caption{Fits (shown in red) of the Curie-von Schweidler equation to LRS and HRS after setting the respective states with voltage sweeps between 0 V and +1 V and -4 V. The extrapolation of these fits was used to estimate the retention time.}
    \label{fig:fig5}
\end{figure}

Stable bipolar resistive switching was present in the as-fabricated device without the requirement of an electroforming process. This indicates that the nature of the switching is not mediated by the formation of local filamentary paths, as is seen in many oxides that are initially insulating. The Schottky nature of the transport characteristics is evident from the temperature dependence of the charge-transport measurements shown in Fig. \ref{fig:fig1}(c). The importance of the existence of the Schottky barrier in realising switching is evident from junctions where no proper Schottky contact was established; these show an absence of hysteresis, even with large voltage sweeps, up to -10 V (Fig. S1 and Fig. S2). Both these observations confirm the interfacial origin of the resistive switching \cite{Yin2017,Yin2015,Chen2011,Dong-Wook2010}. 
\\
\indent Estimations of the barrier heights in the LRS and HRS, realized using a \textit{set} voltage of +1 V and a \textit{reset} voltage of -4 V, were made by fitting Eq. \ref{Eq:thermionic} in the forward bias regime (Fig. S3). This gives a value of 0.72$\pm$0.01 eV for the LRS and 0.76$\pm$0.01 eV for the HRS. The height of the barrier is smaller in the LRS than in the HRS, allowing for an increase in the number of electrons able to overcome the barrier in the forward bias regime, where charge transport is governed by thermionic emission (see Fig. \ref{fig:fig1}(b)). 
\\
\indent Resistive switching is the result of a change in the charge at the interface from a more negative to a more positive value between the HRS and LRS. In literature, both the redistribution of oxygen vacancies and the trapping and de-trapping of electrons at the interface under electric fields have been linked to this change \cite{Mikheev2014,Yin2017,Chen2011,Fan2017}. After switching, it is observed that both the LRS and HRS asymptotically approaches a stable value over time: the resistance of the LRS becomes higher, while the resistance of the HRS becomes lower. In both cases, the resistance $R$ over time $t$ could be fit using the Curie-von Schweidler equation $R\propto t^n$ \cite{Mikheev2014} with rate related exponents, $n$, for the LRS and HRS of 0.685 and -0.244 respectively (see Fig. \ref{fig:fig5}). This change is associated with capacitive charging and is typically observed in high-$\kappa$ materials due to the trapping of charges under bias \cite{Mikheev2014,Yin2015,Yin2017,Ni2007}.
\\ \indent
Under the influence of a reverse bias voltage, the mobile electrons are trapped at defects and oxygen vacancy sites which results in a reduction in the positive charge at the interface. Consequently, the barrier becomes higher and wider resulting in a switch to the HRS. When electrons become de-trapped from the interface, the charge in this region becomes more positive, resulting in a reduction of the height and width of the Schottky barrier giving the LRS. When switching from the HRS to LRS, charges are de-trapped, and more trap states are available at the interface. Over time, negative charges will again become trapped in these states resulting in a progressive increase in the resistance of the LRS. Fig. \ref{fig:fig2}(a) shows that the hysteresis in the forward bias direction (governed by thermionic emission) is less significant than in the reverse direction (governed by tunneling) which suggests that the resistive switching does not originate purely from changes in the profile of the barrier. 

\indent Trap-assisted tunneling may also contribute to transport and when the trap states become filled, these paths become blocked so that tunneling is suppressed \cite{Fan2017}. When the magnitude of the reverse bias \textit{reset} voltage is increased, the magnitude of the electric field across the interface increases and electrons going from the metal to the semiconductor side tunnel from states further below the Fermi level. Consequently, a larger range of interfacial states is available in which they can become trapped, resulting in an increase in the resistance of the HRS obtained upon increasing the \textit{reset} voltage.
\\ \indent
Fig. \ref{fig:fig4} demonstrates that there is very little variation in the HRS after repeat cycles, and there is no difference in the value realized by the sweeps and pulses. The LRS state, on the other hand, is affected more significantly and the resistance achieved with a longer setting process is noticeably lower and reproducible upon cycling. These observations are in accordance with the increasing resistance of the LRS upon cycling (Fig. \ref{fig:fig3}) and indicate that the setting process is slower than the resetting. While, within the time the states were monitored, the difference in the resistance states remained substantial, the progressive change of the LRS (HRS) towards a state of higher (lower) resistance could indicate that over time both will tend to a steady state of intermediate resistance. By extrapolating the Curie-von Schweidler fits from Fig. \ref{fig:fig5}, an estimate of 12,400 s ($\sim$ 3.5 hours) is obtained for this retention time. A similar analysis was done for the measurement performed using pulses to \textit{set} the states and here a retention time of 4150 s ($\sim$ 1 hour) is estimated (Fig. S4). This kind of relaxation is typically observed in systems where the switching mechanism is related to charge trapping and de-trapping \cite{Fan2017}.\\ \indent
Fig. \ref{fig:fig4} eludes to a temporal dependence on the setting process to realize the LRS. The two different methods for setting the LRS can be used to mimic learning and forgetting in short-term memory: when a relatively long stimulus (a positive bias voltage) is used to \textit{set} the device to a state of low resistance and the resistance is monitored immediately after, there is a significant decrease in resistance with respect to before the stimulus. Over time, this state will approach a state more similar to before the stimulus \cite{Yang2012,Yin2017}. If, on the other hand, a shorter pulse is used to \textit{set} the LRS, the state of the device is less significantly affected. Like before, directly after presenting the stimulus the resistance state will be lowest, but in this case, the device will tend to a state reminiscent of its state before the stimulus much faster. This can be compared to the fading of memory as time passes: less significant events (in this case, a shorter pulse) will be forgotten faster.
\begin{figure}[H]
    \centering
    \includegraphics[scale=0.35]{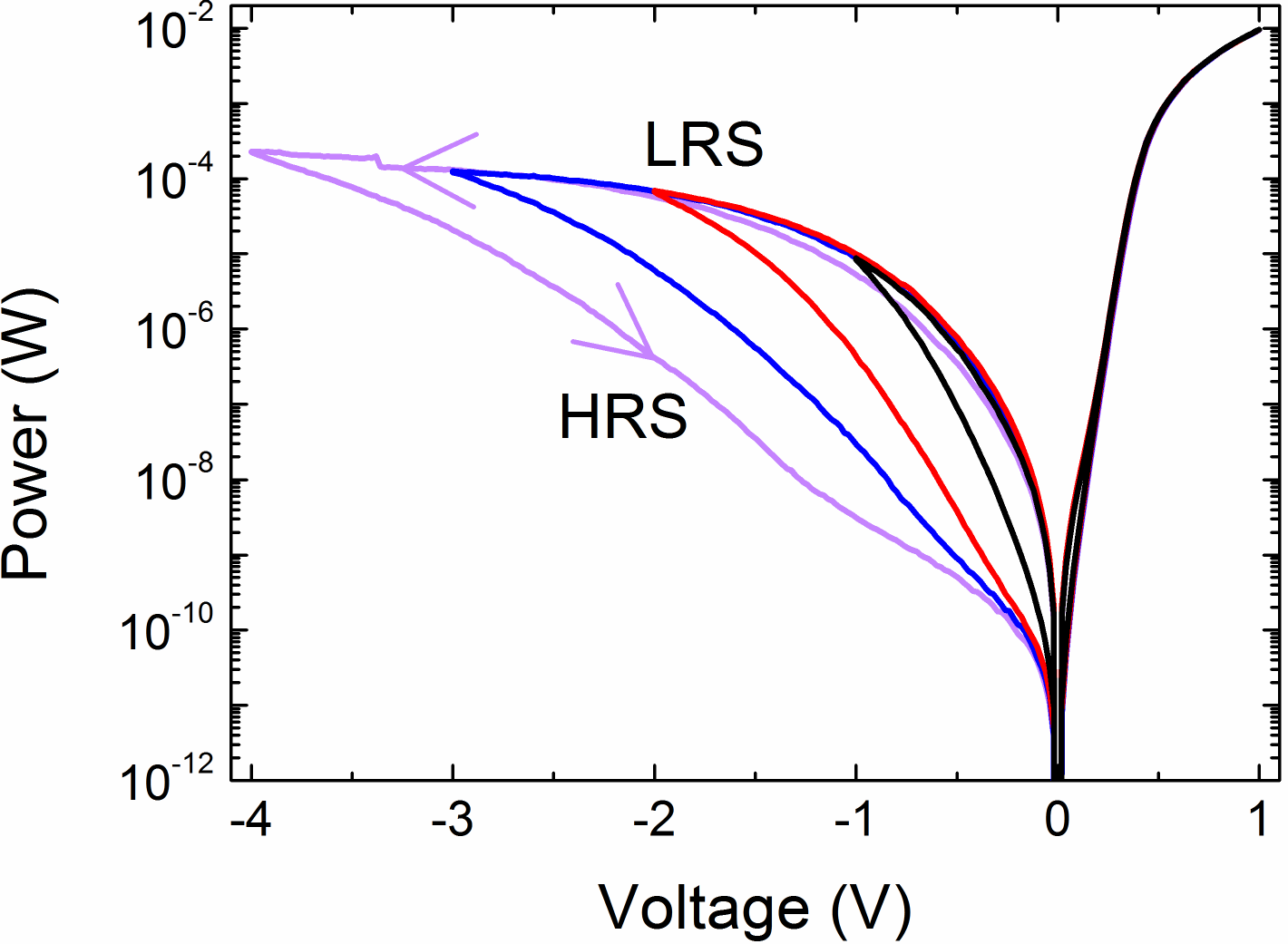}
    \caption{Estimation of power consumption during series of I-V cycles shown in Fig. \ref{fig:fig2}(a).}
    \label{fig:fig6}
\end{figure}
 \indent
From Fig. \ref{fig:fig2}(a) it is clear that multi-level switching can readily be achieved by changing the magnitude of the \textit{reset} voltage. In this way, a plethora of resistance states can be defined at a single low bias read voltage. The gradual, rather than digital, nature of the resistance change not only mimics the analog nature of synapses, but also allows for a large storage. The small amount of current flowing in the reverse bias regime gives rise to very large resistance values (plotted in Fig. \ref{fig:fig2}(b)) both in the LRS ($\sim$ 10$^5$ $\Omega$) and in the HRSs (up to $\sim$ 10$^8$ $\Omega$). In addition to high resistance values, the resistance ratios between the HRS and LRS are also large (see Fig. \ref{fig:fig2}(c)): using -1 V as a read voltage, a maximum ratio of $\sim$ 10$^3$ can be obtained. 
The power ($P$) consumption of this device is shown in Fig. \ref{fig:fig6} and estimated from every point of the I-V plot in Fig. \ref{fig:fig2}(a) using $P=IV$. From the graph it is seen that using a read voltage of -1 V, between 10$^{-9}$ W and 10$^{-5}$ W is required to read four distinct states, when they have been `trained' in different cycles. 
\section{Conclusion}
Bipolar resistive switching at the Schottky interface of Ni and Nb-doped SrTiO$_3$ has been investigated. Measurements have been conducted to specifically address characteristics important for neuromorphic computing, including endurance upon repeat cycles, retention, power consumption and multi-level switching. By changing the \textit{reset} voltage multi-level switching was demonstrated between highly resistive states with resistance variations up to three orders of magnitude. Read powers in the nW regime were realized with junctions of relatively large area. The power consumption of a junction is expected to decrease when it is scaled down to the nanometer regime. The resistive switching showed no permanent degradation during a sequence of 100 switching cycles and showed no signs of fatigue over subsequent measurements. Time was identified as an important parameter governing the setting of the LRS. The retention time of the LRS was seen to increase upon increasing the time of a presented voltage stimulus in a fashion similar to the processes of learning and forgetting in the human brain. 
\section{Acknowledgements}
The authors would like to thank Dr. Terry Stewart (Computational Neuroscience Research Group at Waterloo Centre for Theoretical Neuroscience), Dr. Srikant Srinivasan and Aijaz Hamid (School of Computing and Electrical Engineering at the Indian Institute of Technology, Mandi) for their helpful insight and discussions. We have benefited from scientific talks and discussions at the CogniGron Centre, University of Groningen. AG, AD and TB acknowledge technical support from J.G. Holstein, H.M. de Roosz, T. Schouten and H. Adema. AG is supported by the Top Master Programme in Nanoscience and AD by the Dieptestrategie grant 2014 from Zernike Institute for Advanced Materials at the University of Groningen.
%\bibliography{main}
%merlin.mbs apsrev4-1.bst 2010-07-25 4.21a (PWD, AO, DPC) hacked
%Control: key (0)
%Control: author (8) initials jnrlst
%Control: editor formatted (1) identically to author
%Control: production of article title (-1) disabled
%Control: page (0) single
%Control: year (1) truncated
%Control: production of eprint (0) enabled
%

\end{document}